\documentclass[%
longbibliography,
 reprint,
superscriptaddress,
 amsmath,amssymb,
 aps,
]{revtex4-1}
\usepackage{float}
\usepackage[english]{babel}
\usepackage{comment}
\usepackage{natbib}
\usepackage{ulem}
\usepackage{graphicx}% Include figure files
\usepackage{dcolumn}% Align table columns on decimal point
\usepackage{bm}% bold math
\usepackage{xcolor}
\usepackage{hyperref}
\DeclareUnicodeCharacter{2009}{\,} 
\DeclareUnicodeCharacter{2212}{-}
\begin{document}

\preprint{APS/123-QED}

\title{Self focusing Hybrid Skyrmions in spatially varying canted {ferro}magnetic systems}% Force line breaks with \\

\author{Hamed Vakili}
\email{hv8rf@virginia.edu}
\affiliation{
Department of Physics,University of Virginia, Charlottesville, VA 22903 USA
}
\author{Yunkun Xie}%
\affiliation{School of Electrical and Computer Engineering,
University of Virginia, Charlottesville, VA 22903 USA}
\author{Avik W. Ghosh}%
\affiliation{
Department of Physics,University of Virginia, Charlottesville, VA 22903 USA
}
\affiliation{School of Electrical and Computer Engineering,
University of Virginia, Charlottesville, VA 22903 USA}

\date{\today}% It is always \today, today,

\begin{abstract}
Magnetic Skyrmions are quasiparticle configurations in a magnetic film that can act as information carrying bits for ultrasmall, all electronic nonvolatile memory. The skyrmions can be nucleated and driven by spin orbit torque from a current driven in a heavy metal underlayer. Along its gyrotropic path, a Magnus force can cause a skyrmion to be annihilated at the boundaries. By combining interfacial and bulk Dzyaloshinskii-Moriya interactions (DMI), for instance by using a B20 material on top of a heavy metal (HM) layer with high spin-orbit coupling, it is possible to engineer a hybrid skyrmion that will travel parallel to the racetrack with zero Magnus force. We show that by using a spatially varying interfacial DMI, a hybrid skyrmion will automatically self-focus onto such a track as its domain angle evolves along the path. Furthermore, using a gate driven voltage controlled magnetic anisotropy (VCMA), we can control the trajectory of the hybrid skyrmion and its eventual convergence path and lane selection in a racetrack geometry.
\end{abstract}

\pacs{Valid PACS appear here}
\maketitle
\begin{figure}[h]
\includegraphics[width=\linewidth]{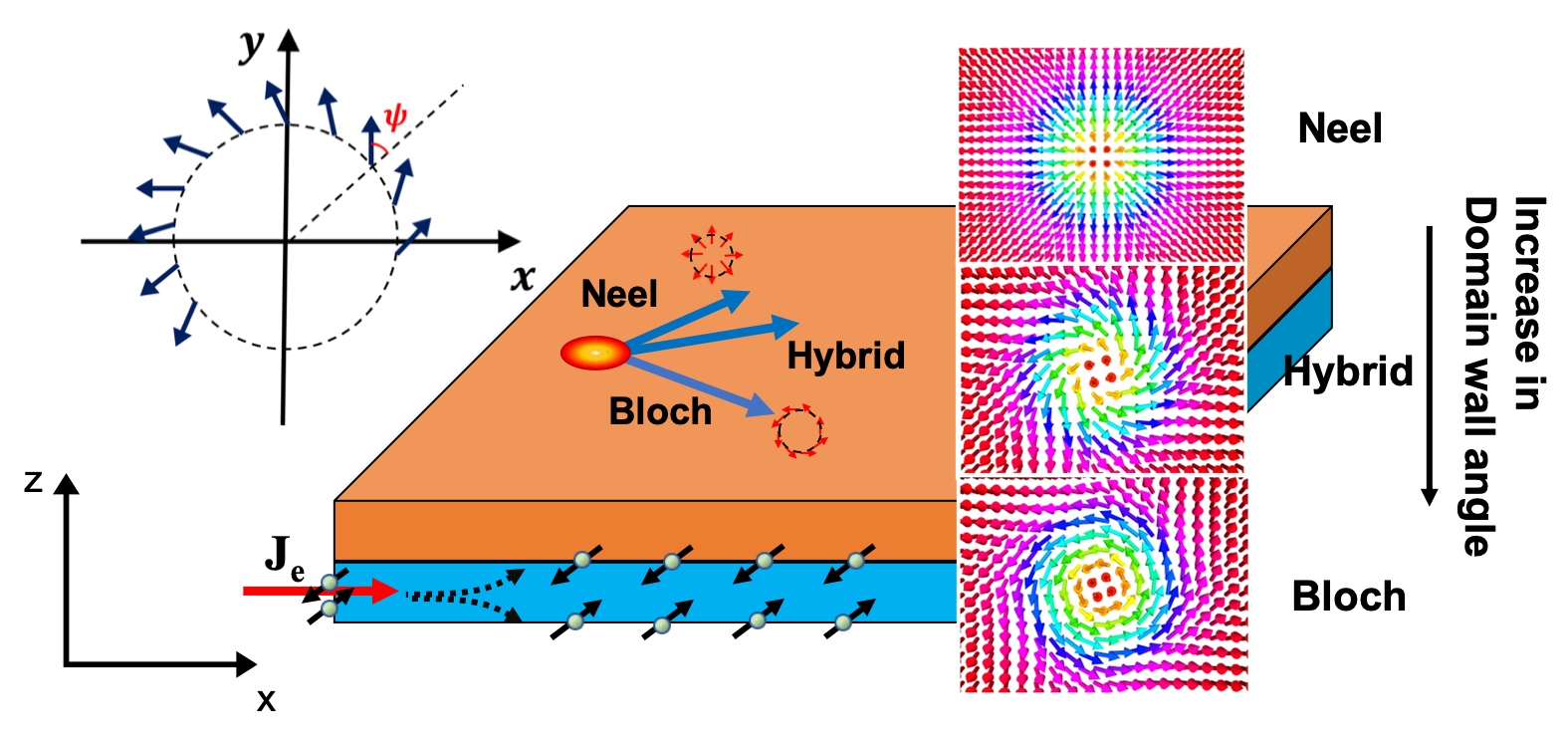}
\caption{Schematic view of a skyrmion in a ferro/ferrimagnetic material (orange). In a FM/HM heterojunction, an applied current in a HM underlayer (black) generates a spin Hall effect that separates opposite spins,  resulting in  spin  injection and  torque applied in the FM layer. For a given current, N\'{e}el and Bloch skyrmions with orthogonal domain wall angle $\psi$ will move perpendicular to each other, so that a suitably engineered hybrid can move along the current direction.}
\label{hybrid}
\end{figure}

 
%\tableofcontents
%\section{Introduction}
\indent Magnetic skyrmions are localized spin textures, potentially much smaller than domain walls. Individual skyrmions can be driven like particles in clean magnetic films, making them interesting candidates for high density information storage such as racetrack memory. Skyrmions are typically nucleated as metastable states on the magnet's energy landscape through a competition between exchange, anisotropy, DMI and stray field energies \cite{bogdanov1989,bogdanov2001,everschor-sitte_perspective_2018,finocchio_promise_2019}. In particular, the DMI energy comes from breaking bulk or interfacial inversion symmetry. The former, bulk DMI (bDMI), exists in chiral magnets such as B20 materials (MnSi, FeGe) and results in Bloch skyrmions with a ninety degree domain angle $\psi$ (Fig.~\ref{hybrid} inset), while the latter interfacial DMI (iDMI) prefers N\'{e}el skyrmions with zero domain angle. These two types of skyrmions are topologically equivalent, but their dynamical behavior are different \cite{woo2016observation,seki2012observation}. In particular, they tend to move orthogonal to each other due to a topologically generated Magnus force under the action of a spin orbit torque (SOT) \cite{nagaosa,anisotropy}. Eliminating that Magnus force \cite{mitigation, SAFskm,selffocus} typically requires ferrimagnets at their angular momentum compensation point, or tracks with raised edges - operating thus at specific temperatures or along pre-set tracks. 
In a structure with both types of symmetry breaking however, we get hybrid skyrmions with a velocity aligned between N\'{e}el and Bloch \cite{hybriddmi,Hybridmagskm} (Fig.~\ref{hybrid}). For a specific domain angle {the net Magnus and driving force from SOT will be in the direction of the applied current} and allow the skyrmion to move linearly along the current path - but that requires a precise confluence of parameters setting the ratio of 
bulk and inversion asymmetry contributions. 
\\
\indent \textcolor{black}{A number of methods have been suggested in the skyrmion literature to drive them into rectilinear motion and mitigate skyrmion Hall drift. These include anisotropy engineering with raised edges and/or anisotropy gradients \cite{mitigation} and frustrated ferromagnets with fine tuned anisotropy engineering \cite{frustrated}, magnetization compensation using ferrimagnets and synthetic antiferromagnets \cite{selffocus,SAFskm}, and iDMI engineering using material stacks with different iDMI signs  \cite{varyingdmiforce} and hybrid skyrmions with uniform iDMI \cite{Hybridmagskm}. Notably
all these structures are spatially uniform along the transport direction. They need precise fabrication to achieve compensation within a single racetrack right from the injection point. All except reference \cite{frustrated} allow no selectivity of multiple racetracks and none allow self-selection into a desired track through dynamic compensation. Such a self-selection can be quite critical to the initialization process in a logic operation \cite{mux,logic}, where pre-existing skyrmions from a common repository can be driven into respective racetracks without the high energy cost and reliability issues of on-site on-demand nucleation. }

\begin{figure}[h]
\includegraphics[width=\linewidth]{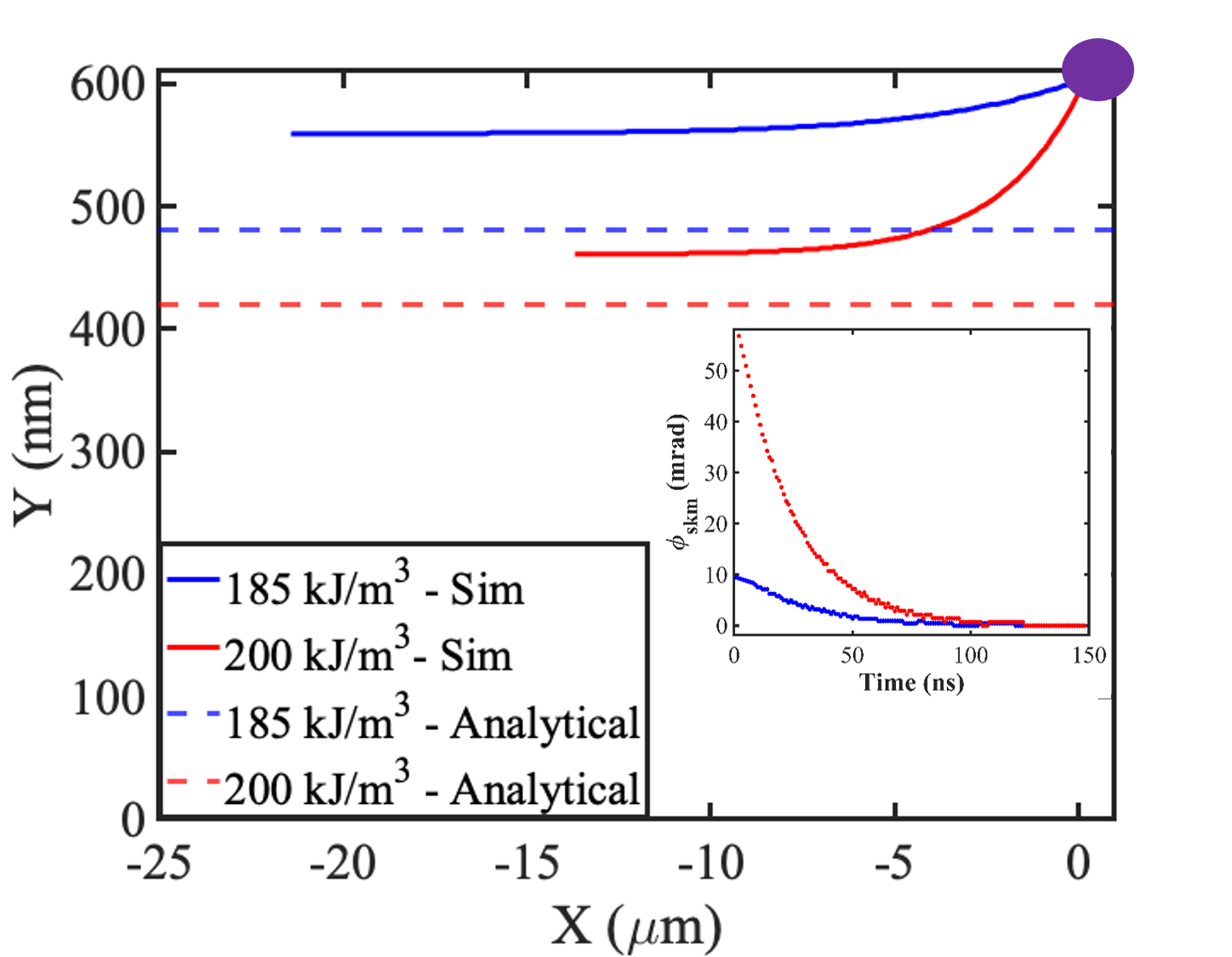}
\caption{Converging lanes for different anisotropy values, simulated (solid) and quasi-analytical (dashed). Increasing anisotropy decreases size $\rho$, causing the domain angle $\psi_c$ to increase according to Eq.~\ref{psicfit}, due to increasing ${bDMI}/{iDMI}$. Here iDMI decreases linearly from top (y = 800 nm, $D_{int} = 0.5 mJ/m^2$) to bottom (y = 0 nm, $D_{int} = 0$). Dashed lines are quasi-analytical predictions of self-converging lanes. Inset shows $\phi_{skm}$ vs time. The purple circle shows the starting point of the skyrmions. }
\label{hybrid3}
\end{figure}

\begin{figure}[h]
\includegraphics[width=\linewidth]{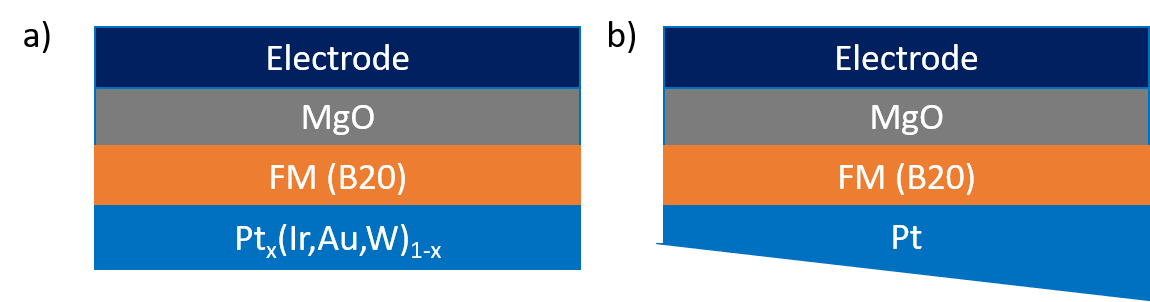}
\caption{Possible stacking options a) Using $Pt_x(Ir,Au,W)_{1-x}$ by varying x value from left to right we can achieve a spatially varying interfacial DMI. b) By changing Pt thickness, \textcolor{black}{or equivalently a varying thickness MgO underlayer between Pt and FM}. The DMI from Pt increases for thicknesses of 1-3 nm and saturates thereafter, and can be used to get a spatially varying interfacial DMI. \textcolor{black}{The top MgO-electrode stack sits on a select part of the racetrack for applying a Voltage Controlled Magnetic Anisotropy (VCMA, Fig.~\ref{10-8}).} }
\label{stacks}
\end{figure}

\begin{figure*}[t]
\includegraphics[width=\textwidth]{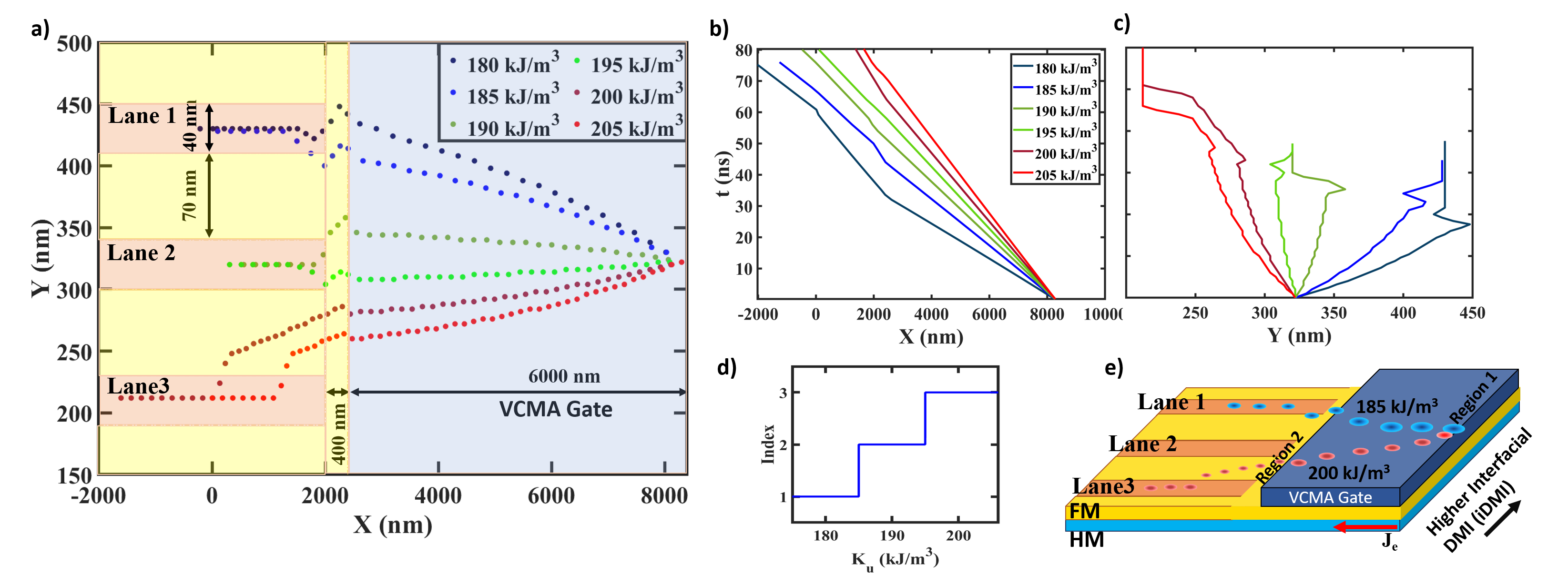}
\caption{{a) Skyrmion self-focusing with iDMI gradient of 312.5 $J/m^3$, varying from 0.45 $mJ/m^2$ at  top, y=800 nm to 0.2 $mJ/m^3$  at bottom, $y=0$. The anisotropy values in the legend are for the VCMA gated region (black). This should be achievable  using an FM $Pt_xW_{1-x}$ underlayer with x varying linearly from top to bottom. Lanes (Orange rectangles) are 40 nm wide each and 70 nm apart, with anisotropy of 190 ${kJ}/{m^3}$. The yellow part is set to have anisotropy of 220 ${kJ}/{m^3}$ so that it will provide the necessary repulsion for skyrmions to stay in lane. Converging lane for a skyrmion with $K_u = 220$ anisotropy is $y = ~150 nm$ (not shown) b),c) x and y vs time. d) The lane index for the final skyrmion convergence vs anisotropy in the VCMA gated region. e) Schematic view of the simulation geometry, black circles are for skyrmions for $K_u = 185 kJ/m^3$ and red circles for  $K_u = 200 kJ/m^3$. Circle sizes correspond to actual skyrmion size}.
}
\label{10-8}
\end{figure*}

\indent In this paper, we show that hybrid skyrmions \cite{randeria} can be made to naturally self-focus along a racetrack with spatially varying DMI (Fig.~\ref{hybrid3}), for instance when it sits on a HM with varying thickness or composition (e.g. Pt$_x$W$_{1-x}$, Fig.~\ref{stacks}). Using analytical results and numerical simulations describing  skyrmion movement, we show that the domain angle for a traveling skyrmion will keep changing with varying DMI until it reaches the cancellation point for the Magnus force. Furthermore, we can dynamically control the converging lane of the skyrmion (Fig.~\ref{10-8}) through a voltage controlled magnetic anisotropy (VCMA) at its interface with a top oxide layer. We can thus gate control the skyrmion trajectory to resonate into specific lanes, while allowing it to diffuse along between resonances. 
\\
%\section{Results}
{\it{Dynamics of hybrid skyrmions.}} One way to move a skyrmion is to use a FM/HM structure. A current in the HM layer, say Pt, separates spins through a spin Hall effect, resulting in the injection of a perpendicular spin current into the FM that then diffuses away from the FM/HM interface. The injected spins precess incoherently around the FM magnetization, applying in the process a spin orbit torque that flips the background spins and drives the skyrmions \cite{racedesign}. The polarization direction of the spin current depends on the symmetry of the HM/FM stack. In our example, the polarization direction is $\hat{P}=\pm \hat{j}\times \hat{n}$, where $\hat{j}$ is the current direction and $\hat{n}$ is interface normal vector, assumed $\parallel \hat{z}$. 

In a FM/HM structure (Fig.~\ref{hybrid}) the driving force from current induced SOT on a N\'{e}el skyrmion in the FM layer will be in the $\hat{x}$ direction, and for a Bloch skyrmion in the -$\hat{y}$ direction. The Magnus force for a N\'{e}el skyrmion will be in the $\hat{y}$ direction and Bloch skyrmion in the $\hat{x}$ direction. The net force from the SOT and Magnus effect generate perpendicular motion of N\'{e}el and Bloch skyrmions, with a hybrid skyrmion moving in between. We see this from the solution of the Thiele equation for the skyrmion velocity  $\boldsymbol{v}$ for a given domain angle $\psi$ \cite{Thiele,skmhallobs}
\begin{align}\label{thiele}
\boldsymbol{F}+\boldsymbol{G}\times \boldsymbol{v}-\mathcal{D}.(\alpha\boldsymbol{v})+\pi B
\theta_{SH}R(\psi)\boldsymbol{j}_{hm} =0 
\end{align}
where $\boldsymbol{F}$ includes unaccounted forces like gradients of effective fields or skyrmion-skyrmion interactions, $\boldsymbol{G} = (0,0,-4\pi N_{sk} )$ is the gyrotropic vector, $N_{sk} = \pm 1$ is the skyrmion winding number, 
$\mathcal{D}$ is the dissipation tensor (assumed isotropic with diagonals ${\cal{D}}_{xx}$), $\alpha$ is Gilbert damping, $B=\pi\hbar I_d\gamma /{2eM_st_{FM}}$ is the SOT pre-factor, $e$ is the electron charge, $\hbar$ reduced Planck's constant, $\gamma$ is the gyromagnetic ratio,  $t_{FM}$ is the FM thickness, $M_s$ saturation magnetization, $I_d = \int dr(r\partial_r \theta +\cos{\theta}\sin{\theta})$, $\theta$ is the angle between magnetic moment $\boldsymbol{m}$ and $z$ axis,  $\rho={R_{skm}}/{\Delta}$ is the size of the skyrmion core $R_{skm}$ relative to the domain wall width $\Delta$. $R(\psi)=\begin{pmatrix} 
\cos{\psi} & \sin{\psi} \\
-\sin{\psi} & \cos{\psi} 
\end{pmatrix}$ is the 2D rotation matrix involving the domain angle $\psi$,    $\theta_{SH}$ is spin hall angle, and $\boldsymbol{j}_{hm}$ is the current density in the HM layer, assumed to be along the $x$ direction.
Solving for $\boldsymbol{v}$:
%\begin{eqnarray}
    %&&v_x=\Big(\frac{G4\pi %B\:}{G^2+\alpha^2\mathcal{D}_{xx}^2%}\:\sin{\psi}+\frac{\alpha 4\pi %\mathcal{D}_{xx}\:B}{G^2+\alpha^2\: %\mathcal{D}_{xx}^2}\: %\cos{\psi}\Big)\theta_{SH}j_{hm}+\n%onumber\\ &&\frac{\alpha %D_{xx}F_x-G %F_y}{G^2+\alpha^2D_{xx}^2}\nonumber%\\
     %&&v_y=\Big(\frac{G4\pi B\: %}{G^2+\alpha^2\mathcal{D}_{xx}^2}\%: \cos{\psi}-\frac{\alpha 4\pi %\mathcal{D}_{xx}\:B}{G^2+\alpha^2\%: \mathcal{D}_{xx}^2}\: %\sin{\psi}\Big)\theta_{SH}j_{hm}+\%nonumber\\ &&\frac{\alpha %D_{xx}F_y+G %F_x}{G^2+\alpha^2D_{xx}^2}
%%\end{eqnarray}
\begin{equation}
\boldsymbol{v} = \displaystyle M^{-1}\Bigl(\boldsymbol{F} - 4\pi B\theta_{SH}R(\psi)\boldsymbol{j}_{hm}\Bigr), ~~~ M = \alpha {\cal{D}} + {\cal{G}} 
\end{equation}
where ${\cal{G}} = 4\pi\left(\begin{array}{cc}0 & N_{sk} \\ - N_{sk} & 0\end{array}\right)$ is the gyrotropic tensor.
From this equation, we can extract the skyrmion hall angle $\phi_{skm} = \tan^{-1}(v_y/v_x)$, and the critical domain angle $\psi_{c}$ where the Magnus force vanishes ($\phi_{skm} = 0$). At zero force $F=0$, we get
\begin{align}\label{skmhall}
    \phi_{skm} &= tan^{-1}\Biggl(\frac{G \; \cos{\psi} - \alpha \: \mathcal{D}_{xx} \;\sin{\psi}}{G\; \sin{\psi}+\alpha  \:\mathcal{D}_{xx}\; \cos{\psi}}\Biggr)\\
    \psi_{c} &= tan^{-1}\Bigl({G}/{\alpha \mathcal{D}_{xx}}\Bigr)~~~{\rm{when~}\phi_{skm} = 0}\label{psic}
\end{align}
with $G = 4\pi N_{sk}$.
%\begin{widetext}
%\textcolor{magenta}
%{
%with non zero Fy:
%\begin{align}\label{skmhall2}
    %\phi_{skm} &= tan^{-1}\Biggl(\frac{(G \; \cos{\psi} - %\alpha \: \mathcal{D}_{xx} \;\sin{\psi})4\pi B \theta %j_{hm}+\alpha \mathcal{D}_{xx} F_y +GF_x}{(G\; %\sin{\psi}+\alpha  \:\mathcal{D}_{xx}\; \cos{\psi})4\pi B %\theta j_{hm}+\alpha \mathcal{D}_{xx} F_x - G %F_y}\Biggr)\\
%\end{align}
%}
%\end{widetext}
The domain angle $\psi$, {set by the ratio of bulk to interfacial DMI}, determines if a skyrmion is N\'{e}el (Figs.~2,~3) ($\psi=0,\pi$), Bloch ($\psi = \frac{\pi}{2},\frac{3\pi}{2}$)  or hybrid ($\psi \neq 0,\frac{\pi}{2},\pi,\frac{3\pi}{2})$.
For a given $R_{skm}$ and $\Delta$, the velocities of Bloch and N\'{e}el skyrmions are perpendicular to each other ($\boldsymbol{v}_{Bloch} . \boldsymbol{v}_{N\acute{e}el} = 0$), so that a proper hybrid skyrmion at domain angle $\psi_c$ moves along the current direction. However, reaching this domain wall angle requires a precise tuning of parameters. 

{\it{Spatially varying iDMI for self-focusing skyrmions.}} Let us now consider a hybrid skyrmion moving in a FM layer with both bulk DMI (e.g. a B20 material like FeGe or MnSi) and a linearly varying interfacial DMI from HM spin-orbit coupling \cite{sheobs}
\begin{align}\label{dmi}
    D_{int}&= D_0 + \lambda y 
\end{align}
The DMI energy density $\epsilon_{DMI} = D_{int} \cos{\psi}-D_{bulk}\sin{\psi}$. 
 {Solving ${\partial \epsilon_{DMI}}/{\partial \psi} = 0$}, we can then get the evolution of $\psi$ as a function of y from Eq \ref{dmi}
\begin{align}\label{psieq}
    \psi(y) = \tan^{-1}\Bigl(\frac{-D_{bulk}}{D_0 + \lambda y}\Bigr)
\end{align}
meaning the skyrmions pick up increasingly more N\'{e}el characteristic during transit. 
The corresponding force from the linearly varying interfacial DMI \cite{varyingdmiforce}: 
\begin{equation}
\boldsymbol{F}_{DMI} = -\frac{\partial E_{DMI}}{\partial \boldsymbol{r}} = 2\pi\lambda\frac{\gamma}{M_s}\Delta I_d \;\cos{\psi}\; \hat{y}
\end{equation}
\textcolor{black}{Where $E_{DMI} = \int 2\pi\epsilon_{DMI}(r\partial_r \theta +\cos{\theta}\sin{\theta})~ dxdy $.} Using a 2$\pi$ domain wall model for the  azimuthal angle $\theta(r)$ (distinct from domain wall angle $\psi$), $\theta(r) = 2\tan^{-1}[\sinh{\rho}/\sinh{(r/\Delta)}]$, the DMI integral $I_d$ can be approximated as $I_d \approx \pi \rho\Delta $. Putting in $\boldsymbol{F}_{DMI}$ in the Thiele equation, we get a correction to the critical angle: $\tan{\psi_c} = (1+C)G/\alpha{\cal{D}}_{xx}$, with $C = 4\lambda et_{FM}\alpha \mathcal{D}_{xx}/\pi\hbar N_{sk}G\theta_{SH}j_{hm}$. As Fig.~\ref{hybrid3} shows, the convergent lane depends on the anisotropy parameter $K_u$, which in turn affects the skyrmion size $\rho$ and thereby the dissipation tensor ${\cal{D}}_{xx}$ responsible for the critical domain angle $\psi_{c}$. The dissipation integral $\mathcal{D}_{xx}$ is defined as \cite{Beachisolated}:
\begin{eqnarray}\label{dxx}
    &&\mathcal{D}_{xx}(\rho) = \int dxdy\;(\partial_x \boldsymbol{m})^2\\
    &&= 2\pi \sinh^2{\rho}\int_0^\infty \displaystyle\frac{[\cosh{2r}+1]r + [\cosh{2r}-1]/r}{\Bigl(\sinh^2{r}+\sinh^2{\rho}\Bigr)^2}dr\nonumber
\end{eqnarray}
in a 2$\pi$ domain wall model. The integral above does not have an analytical solution to our knowledge. However, a simple fit that works quite well for $\rho \sim 0.5-4$ is $\mathcal{D}_{xx} \approx 5.96 \sqrt{4.285+\rho ^2}$, so that:%and $\mathcal{D}_{xx} \approx 2\pi (\rho+1/\rho)$ for $\rho > 4$
\begin{align}\label{psicfit}
\psi_{c} \approx tan^{-1}\Big(\frac{G}{ \; 5.96~\alpha \sqrt{4.285+\rho ^2}}(1+C)\Big)
\end{align}
The anisotropy dependence of the skyrmion size $\rho$ can be seen from an evaluation of the energy integrals within a $2\pi$ model \cite{rohartskyrmion,xie_computational_2019}, and can be approximated as (with zero external magnetic field)
\begin{equation}
\rho (y) = \Biggl(\frac{D(y)}{D_c}\Biggr)^2\frac{C_1}{\sqrt{1-C_2(D(y)/D_c)^4}}
\label{small_skm_R}
\end{equation} 
where $D_c = 4\sqrt{A_{ex}K_u}/\pi$, $C_1 \approx 12.7$, $C_2 \approx 1.06$ and \textcolor{black}{$D(y) = D_{int}(y)\cos{\psi}-D_{bulk}\sin{\psi}$} is the total DMI. Since $D_{int}$ is changing with y (Eq.~\ref{dmi}), $\rho$ is a function of y for a given anisotropy. Solving $\psi_c = \tan^{-1}\Bigl(\frac{-D_{bulk}}{D_0 + \lambda y}\Bigr)$ for y, we get the convergence lane (Fig~\ref{hybrid3}).
%\begin{equation}
%    y = \frac{-1}{\lambda} \Bigl(\frac{D_{bulk}5.96~\alpha \sqrt{4.285+\rho(y) ^2}}{G(1+C)}+D_0\Bigr) 
%\end{equation}
%Solving it for y, we get the converging lane  (Fig~\ref{hybrid3}).
\\
We perform micromagnetic (MuMax3 \cite{mumax}) simulations to verify our assumptions and see skyrmion behavior in a linearly varying iDMI with different anisotropies (Fig.~\ref{hybrid3}).
The parameters used in simulation are taken from \cite{Hybridmagskm,hybriddmi, B20thin,B20param} for FeGe, $M_s = 300$ kA/m, $A_{ex}$ = 7.5e-12 $J/m$, $D_{bulk}$ = 1 $mJ/m^2$, $\alpha = 0.2$, $K_u$ = 185, 200 kJ/m$^3$ (assuming interfacial effects for ultra thin FeGe), $\theta_{SH} = 0.15$ and $j_{hm} = 10^{12} A/m^2$. The FM layer thickness $t_{FM}$ is taken to be 2 nm. \indent We apply a spin current of polarization $ +\hat{y} $ in the $\hat{z}$ direction, arising for instance from SOT in the HM layer from a charge current in the $\hat{x}$ direction. The Magnus force will cause the skyrmion to have a velocity component in the $-\hat{y}$ direction. As the skyrmion comes down the FM, the interfacial DMI decreases, which leads to an increasing domain wall angle $\psi$ (Eq.~\ref{psieq}), until it reaches the critical angle $\psi_c$ where it no longer has a velocity component along $\hat{y}$ and will have a rectilinear motion, self-focusing into a lane in the process. 

Fig.~\ref{hybrid3}a shows the tracks for a skyrmion injected at the low iDMI end, as obtained from a numerical simulation of the Landau-Lifschitz-Gilbert equation (which goes beyond the rigid skyrmion Thiele approximation), for various bulk anisotropy values $K_u$. The horizontal dashed lines show the convergent path coordinates $y$ obtained from analytical approximations \cite{analytical}. The decrease in skyrmion size $\rho$ along the travel path $y$ means that smaller skyrmions (e.g. with higher anisotropy) travel further down before self-focusing. 

\label{eq:small_skm_R}
\indent To get a linearly varying iDMI we suggest two possible approaches (Fig \ref{stacks}) - one is to use a varying thickness of \textcolor{black}{HM layer, or more realistically of an} MgO  film between the HM layer and FM layer, to tune the iDMI. According to the experiments done in \cite{dmithick,mgotuning}, by putting a MgO layer between HM and FM, the iDMI can be increased substantially and this effect increases with thickness of MgO until it saturates for a thickness of around 2 nm. \textcolor{black}{As the MgO layer has significantly higher resistance than the HM layer, most of current will travel trough the HM layer. However, further investigation is needed to understand how the separation between HM and FM layer due to the MgO layer alters the effective momentum transfer.} A second approach is to use a non uniform composition of HM, e.g. $Pt_x(Ir,Au,W)_{1-x}$ \cite{hmcomp, tunecomp}. It has been seen by using a composition of $Pt_x(Ir,Au,W)_{1-x}$ that the iDMI increases with increasing $x$ from 0 to 1. The main fabrication challenge is growing high quality B20 thin films on heavy metals and underlying the MgO.

{\it{VCMA for gate dependent lane selection.}}\indent From Eq.~\ref{psicfit}, if we increase $\rho$, $\psi_{c}$ would decrease. One way to increase $\rho$ is by \textcolor{black}{reducing} the anisotropy $K_u$ and thus $D_c$ (Eq.~\ref{small_skm_R}). \textcolor{black}{We can gate control the skyrmion size through Voltage Controlled Magnetic Anisotropy (VCMA), in effect, voltage gating the Stark shift of the bond between the magnet's $d_z^2$ orbital and the oxygen $p_z$ orbital of an oxide layer like MgO grown on the side of the sample opposite the HM layer (Fig.~\ref{stacks}). The equation describing the perpendicular magnetic anisotropy variation with gate voltage in a VCMA set-up is
\begin{align}\label{vcma}
    K_u(V) = K_u (0) -{\xi V}/{t_{ox}t_{FM}} 
\end{align}
with $t_{ox}$ the oxide thickness and $\xi$ the VCMA coefficient. This means with a gate voltage, we can change the anisotropy and $D_c$, the size $\rho$ of the skyrmion, the dissipation tensor ${\cal{D}}_{xx}$ and ultimately the critical angle $\psi_c$ - making the skyrmion converge to a different lane.} Assuming a thickness of 1 nm for oxide and FM layers, we would need a $\xi = 50\; fJ/Vm$ to get a $10^5\;J/m^3$  change in anisotropy per 1V.  $\xi$ of 50-100  fJ/Vm has been achieved for CoFeB of 1 nm thickness \cite{cofeb50,cofeb}, whereas by doping FM/oxide interface $\xi$ larger than 100 fJ/Vm has been reported \cite{vcmareview,vcmaadvance}, suggesting that our device parameters are quite realizable for lane control. To get a higher separation of lanes, we can choose a smaller $\lambda$ term  in Eq \ref{dmi}. This can be done by making the doping $x$ variation slower in Fig.~\ref{stacks}a, or using Pt with a slower thickness variation, Fig ~\ref{stacks}b.
\\
\indent Fig.~\ref{10-8} shows how skyrmions can be made to self-focus into a set of prefabricated racetracks using a VCMA gate \cite{zhang_control_2018}. Anisotropy in the fabricated lanes is $K_u = 190 kJ/m^3$, separated by regions with higher magnetic anisotropy $K_u = 220 kJ/m^3$. While the skyrmion domain angle evolves during transport along $y$, the VCMA gate changes the local anisotropy and alters the bending of the skyrmion track, attempting to set them into alignment with the racetracks. \textcolor{black}{ The VCMA gate can change the anisotropy of the magnetic film under it, enough to bend the skyrmion path towards one of the lanes. The competition between attractive force from the lanes vs. Magnus force in region 2 determines which lane a
skyrmion will converge to.} When the alignment is not perfect (non-resonant skyrmion), the skyrmions instead enter the region between the lanes. The lane locations and relative anisotropies are designed so the skyrmions entering the regions in between the tracks see a Magnus force in the $-\hat{y}$ direction, as well as an attractive force towards the nearest racetrack. The net force will determine which track the non-resonant skyrmions finally converge to. Once they enter a lane, the repulsive force from the interstitials cancels out the Magnus force and the skyrmions stay in  lane. \textcolor{black}{Since the interstitials have higher anisotropy, even skyrmions which miss the lanes eventually end up in one of the lanes whereupon their combined anisotropy repulsion and Magnus force vanishes again.} As a result, we get a stepfunction-like quantized behaviour (Fig~\ref{10-8}b) for the final $y$ coordinate of the convergent skyrmion lanes as a function of the VCMA engineered anisotropy, implying a robust well controlled scheme for directing skyrmions into the lanes. We emphasize that this ability to control lanes dynamically is unique to skyrmions with flexible domain angles, and is not achievable with domain walls. %In Fig.~\ref{fig:rvstime} we see the evolution of $X,~Y$ versus time. The X position.
%\begin{figure}[H]
%    \centering
%    \includegraphics[width=\linewidth]{rvstime2.png}
%    \caption{Skyrmion position vs time}
%    \label{fig:rvstime}
%\end{figure}

{To see thermal effects, we performed stochastic simulations at room temperature (supplementary, fig S1\cite{Supp}). Thermal effects cause fluctuation in the skyrmion's $y$ coordinate, which may create uncertainties near the separatrix between trajectories going to neighboring lanes. The thermally limited transition, inherent in all devices, can be reduced with materials with higher exchange parameter, or a slower variation of iDMI, $\lambda$ compared to the separation of lanes. As long as the lane separations are much larger than the position fluctuations, the system should operate reliably at room temperature. Overall, as in any other device, we need to design and optimize around such inherent instabilities. }\\
\textcolor{black}{{\it{Summary.}} In this paper we have shown a dynamic method of controlling the skyrmion movement. In the literature, other methods of achieving self convergence has been investigated \cite{selffocus,Hybridmagskm,hybriddmi} but the self convergence is not dynamic or needs extra fine turnings. The advantage of our method is the ability to achieve multiple converging lanes. In addition we can dynamically control position of converging lanes and skyrmion movement. The self convergence would circumvent some of the extra fine tuning needed compared to other methods. By using a local VCMA gate, we showed an added degree of flexibility can be achieved which in turn can be used to dynamically manipulate the logic flow. The existence of multiple convergence lane makes it possible to do more complicated logic operations compared to the racetracks\cite{racedesign,mux,luo_reconfigurable_2018} proposed in the literature. We have also provide a quasi analytical method to predict the convergence lane which matches reasonably well with the simulations results.  }

{\it{Acknowledgments.}} This work is funded by the DARPA Topological Excitations in Electronics (TEE) program (grant D18AP00009). We would like to thank Joseph Poon, Mircea Stan, Andy Kent, Prasanna Balachandran and Samiran Ganguly for  insightful  discussions. 
\bibliography{main}

\begin{thebibliography}{40}
\bibitem [{ana()}]{analytical}%
  \BibitemOpen
  \href@noop {} {\bibinfo  {journal} {We are treating skyrmion as a point
  particle with zero dimensions. Since DMI is linearly varying this assumption
  will result in a small disagreement with LLG simulations.}\ }\BibitemShut
  {NoStop}%
\end{thebibliography}
\end{document}